# Theoretical Insights into Layered Metamaterials with Enhanced Thermal and Mechanical Properties

**Hossein Rokni[1], Patrick Singleton[1], Yuanlong Zheng[1,2], Connor Blake[1], Haoran Lin[1], Shuolong Yang[1]**

[1] *Pritzker School of Molecular Engineering, University of Chicago, Chicago, IL 60637, USA*
[2] *Department of Physics, University of Chicago, Chicago, IL 60637, USA*

## ABSTRACT

The inherent trade-off between ultra-low thermal conductivity and high mechanical rigidity in natural materials limits their utility in advanced applications. Inspired by the unique architecture of layered honeycomb structures, this study introduces a new class of metamaterials designed to overcome these constraints. By systematically exploring unit cell configurations and stacking arrangements, we demonstrate that a zigzag internal geometry, analogous to rhombohedral graphene stacking, optimizes thermal insulation while maintaining relatively high mechanical rigidity. Our finite element simulations predict that these layered structures can achieve a thermal conductivity of 12.5 mW/(*m.K*) using zirconia as the constructing material, theoretically outperforming state-of-the-art ceramic aerogels while maintaining robust mechanical stability. This novel approach paves the way for designing next-generation super-insulating materials with customizable mechanical properties, enabling innovative applications in extreme environments, lightweight aerospace structures, and advanced thermal management systems.

## 1. INTRODUCTION

Metamaterials represent a groundbreaking class of engineered materials, designed to achieve physical properties beyond the capabilities of conventional materials through innovative unit cell architectures. Acoustic metamaterials, for instance, have been engineered for applications such as noise reduction and acoustic cloaking **[1]**, while thermal metamaterials enable advanced heat flow control, including thermal invisibility cloaks and high-efficiency insulators **[2]**. Mechanical metamaterials exhibit extraordinary properties such as a negative Poisson's ratio and tunable stiffness **[3]**. Similarly, terahertz **[4]**, electromagnetic **[5]**, and chiroptical metamaterials **[6]** have revolutionized optical technologies, enhancing device performance and enabling precise polarization control. In addition, energy-absorbing **[7]**, elastic wave **[8]**, optical **[9]**, and biomedical **[10]** metamaterials continue to expand the boundaries of wave manipulation, structural engineering, and multifunctional system design. The growing demand for materials with unique combinations of physical properties, particularly ultra-low thermal conductivity coupled with high mechanical rigidity, underscores the need for further exploration of metamaterials for applications such as advanced electronics, lightweight aerospace structures, and biomedical devices **[11–13]**.

Despite significant advancements, the inherent limitations of natural materials pose challenges in combining ultra-low thermal conductivity with high mechanical rigidity, as these properties are typically mutually exclusive. For instance, ceramic aerogels are well-known for their ultra-low thermal conductivity (< 0.1 W/(*m.K*)) but only have a maximum compressive strength of 2–100 kPa, lacking the mechanical strength required for structural applications **[14]**. On the other hand, materials like diamond possess maximum compressive strengths of 60–130 GPa but are excellent thermal conductors with thermal conductivities in the range of 2000-2500 W/(*m.K*). These contrasting properties highlight the potential of metamaterials with tailored unit cell designs to overcome these limitations and unlock new performance regimes. This potential has driven efforts to explore thermal properties in mechanical metamaterials using conventional unit cell designs, including 3D strut-based configurations and triply periodic minimal surface (TPMS) structures **[15-18]**. Although TPMS-based designs have achieved thermal conductivities as low as 0.25 W/(*m.K*) **[18]**, such conventional designs in general have struggled to achieve the ultra-low thermal conductivity range, typical of ceramic aerogels (<0.1 W/(*m.K*)), due to inherent geometric constraints, limited surface area-to-volume ratios, and inefficient heat path configurations.

To address these limitations, we propose a novel class of layered unit cells, inspired by the structural principles of 2D crystals such as graphene. These layered architectures leverage high surface area-to-volume ratios and tunable interlayer interactions to achieve unprecedented thermal and mechanical performance. Using finite element analysis, we systematically compare the heat conduction and mechanical response of layered unit cells with 3D strut-based and TPMS configurations, demonstrating that the ABC-stacked unit cell yields the lowest effective thermal conductivity ($\kappa_{\mathrm{eff}}$) and the AB-stacked one leads to optimal mechanical rigidity, characterized by the maximum stress ($\sigma_{\max}$). Further, we explore the impact of internal structural parameters, including the number, thickness, and placement of pillars and layers, on the effective thermal conductivity and mechanical behavior at the metamaterial level. Our results reveal that a zigzag configuration within the layered structure optimally balances thermal insulation and mechanical strength. By refining the number and thickness of layers and pillars, we achieve an ultra-low thermal conductivity of 12.5 mW/(*m.K*) using zirconia as the constructing material, theoretically surpassing the best-reported ceramic aerogels **[14]**. These findings establish the potential of engineered layered structures, which can be fabricated using 3D additive manufacturing techniques, as cutting-edge solutions for super-insulating materials with robust mechanical stability, opening new avenues for advanced thermal management and structural applications.

## 2. RESULTS AND DISCUSSIONS

### 2.1. Thermal and Mechanical Simulations at the Unit Cell Level

We initially evaluate the thermal and structural performance of layered metamaterials at the unit cell level, comparing the effective thermal conductivity and mechanical rigidity of layered unit cells, inspired by AA (**Fig. 1a**), AB (**Fig. 1b**), and ABC (**Fig. 1c**) stacking modes, commonly found in 2D crystals, with conventional 3D strut and TPMS unit cells (**Fig. 1d**). Throughout this work we use zirconia to construct metamaterials, utilizing its relatively low thermal conductivity (3 W/(*m.K*)) and high compressive strength (2000 MPa). Finite element simulations are performed on single 2 mm cubic zirconia unit cells with a volume fraction of 32%, sandwiched between solid plates (**Fig. S1**, supporting information). For thermal analysis, the top plate is maintained at 500 °C, and the temperature is calculated at the interface between the unit cell and the bottom plate under combined conductive and radiative heat transfer conditions. Using the temperatures at the hot and cold ends of the unit cell, along with the overall heat flux, we evaluate the effective thermal conductivity, $\kappa_{\mathrm{eff}}$. For mechanical analysis, a 10 N load is applied to the top plate while the bottom plate is fixed, and the resulting maximum stress, $\sigma_{\max}$, within the unit cell is determined. It is important to note that a lower $\sigma_{\max}$ indicates better mechanical rigidity of the unit cell.

The results reveal that the AB and ABC stacking modes of the honeycomb lattices achieve the lowest thermal conductivities while maintaining reasonably low $\sigma_{\max}$, with the performance depending on their specific arrangement. This behavior can be attributed to the distinctive ability of layered structures, regardless of the stacking mode, to confine heat flow within individual layers, effectively acting as heat reservoirs. Notably, the pillar arrangements play a crucial role in determining the amount of heat stored within these layers. For example, the ABC stacking mode, distinguished by its zigzag heat path, maximizes heat storage within each layer, resulting in the lowest $\kappa_{eff}$ among the configurations while preserving reasonable mechanical rigidity. In contrast, the AB stacking mode, characterized by a carefully optimized arrangement of straight pillars and an intermediate effective heat path (shorter than ABC but longer than AA), achieves a well-balanced combination of low thermal conductivity and high mechanical rigidity, making it an excellent choice for applications that demand both superior thermal insulation and robust structural integrity. Note that the effective heat path in a unit cell considers all possible heat transfer routes from the top to the bottom and we calculate the average path length as the effective heat path length.

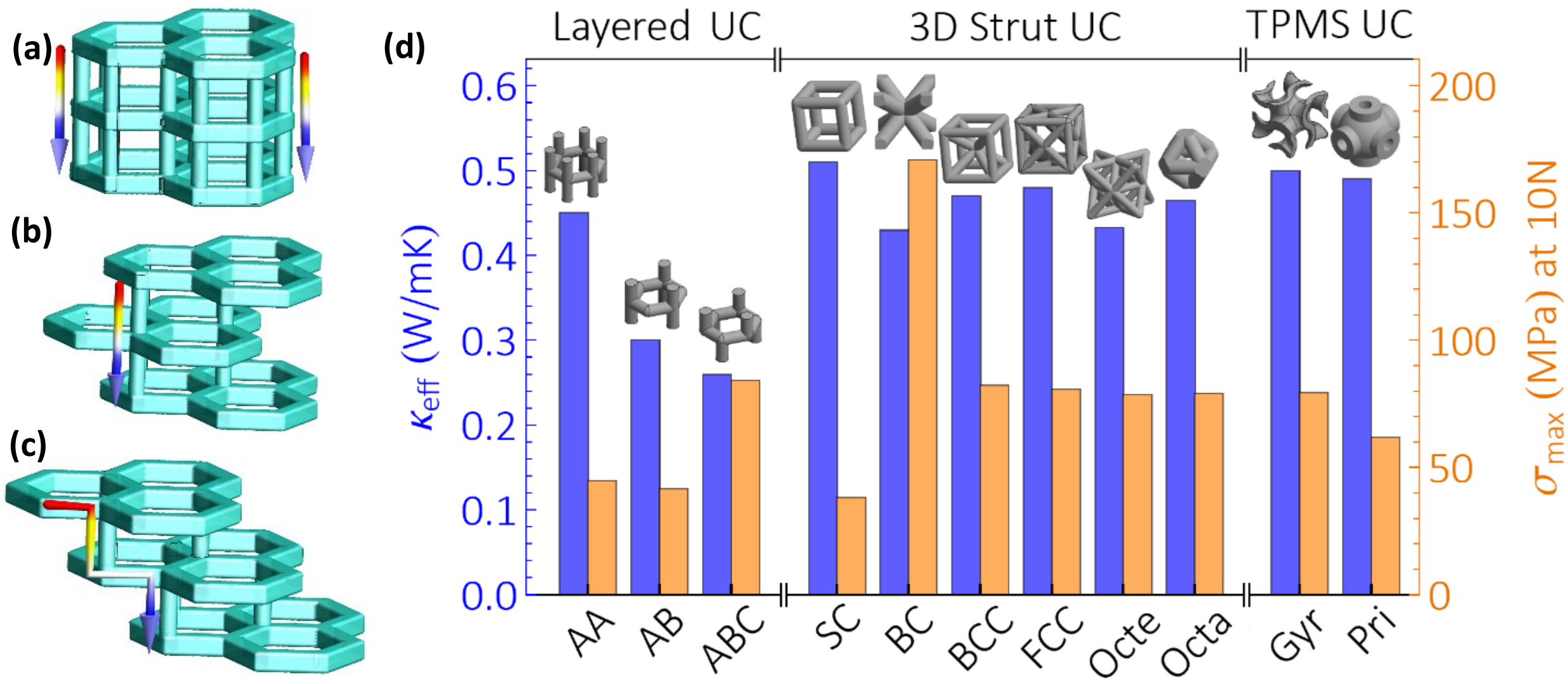

**Fig. 1. Mechanical and thermal properties of layered unit cells.** (a) Straight AA stacking, (b) straight AB stacking, and (c) zigzag ABC stacking, showing distinct conductive heat paths in layered unit cells. (d) Comparative analysis of effective thermal conductivity and maximum stress under a 10 N load for layered unit cells versus conventional unit cells, based on a single unit cell size of 2 mm with a volume fraction of 32%.

### 2.2. Thermal and Mechanical Simulations at the Metamaterial Level

While layered metamaterials can be generated through periodic repetition of the layered unit cells, their reconfigurable nature allows for further optimization of mechanical and thermal properties by engineering the number, thickness, and location of the pillars and layers. To specifically reduce $\kappa_{eff}$, this study goes beyond the periodic structures shown in **Figs. 1(a)-(c)** by reducing the number of pillars between adjacent planes. Notably, this study does not focus on optimizing the size, shape, or arrangement of voids within each layer, as their effect on the in-plane thermal conductivity of ceramic layers has been well-established in previous studies **[19-21]**.

At the metamaterial level, we begin by optimizing pillar locations in a 5-layer system, where 20 pillars are positioned between adjacent layers. Each layer has voids arranged in a triangular lattice, conceptually mimicking the honeycomb structure in **Fig. 1**. Our goal is to minimize a combined metric of thermal conductivity and mechanical stress (**Fig. 2**), defined as follows:

$$\left[\frac{(k_{\mathrm{eff}}/k_0)^2}{2}+\frac{(\sigma_{\max}/\sigma_0)^2}{2}\right]^{1/2} \tag{1}$$

$\kappa_0$ and $\sigma_0$ represent the maximum thermal conductivity and mechanical stress, respectively, identified among over 200 layered structures with different pillar configurations (**Fig. 2**). Using the same boundary conditions as in the unit cell analysis, the topmost layer is maintained at 500 °C and the average temperature and heat flux at the bottommost layer are calculated for thermal analysis. For mechanical analysis, a uniform load of 10 N along the vertical direction is applied to the topmost layer, while the bottommost layer is fixed, and the resulting $\sigma_{\max}$ within the 5-layer system is determined.

Our results show a weak correlation between $\kappa_{\mathrm{eff}}$ and $\sigma_{\max}$ across over 200 pillar configurations (**Fig. S2**, supporting information), yet specific cases should be carefully analyzed. In the "inner-outer" pillar arrangement, where the pillars above each layer are concentrated near the outer edge of the disk and those below are near the inner edge (bottom-right panel in **Fig. 2**), we achieve the lowest thermal conductivity of 0.06 W/(*m.K*) but the highest $\sigma_{\max}$ of 84.4 MPa. This outcome is expected, as the inner-outer arrangement maximizes not only the effective heat path but also the concentrated stress due to the non-uniform pillar distribution between each pair of adjacent layers. In contrast, the "zigzag" pillar configuration at the metamaterial level, where the pillars above each layer are arranged in a uniform triangular lattice and the ones below are arranged in another triangular lattice with a horizontal offset (top-right panel in **Fig. 2**), we achieve an optimal balance between thermal conductivity and mechanical rigidity, characterized by a combined metric of 0.63, as defined in **Eq. (1)**.

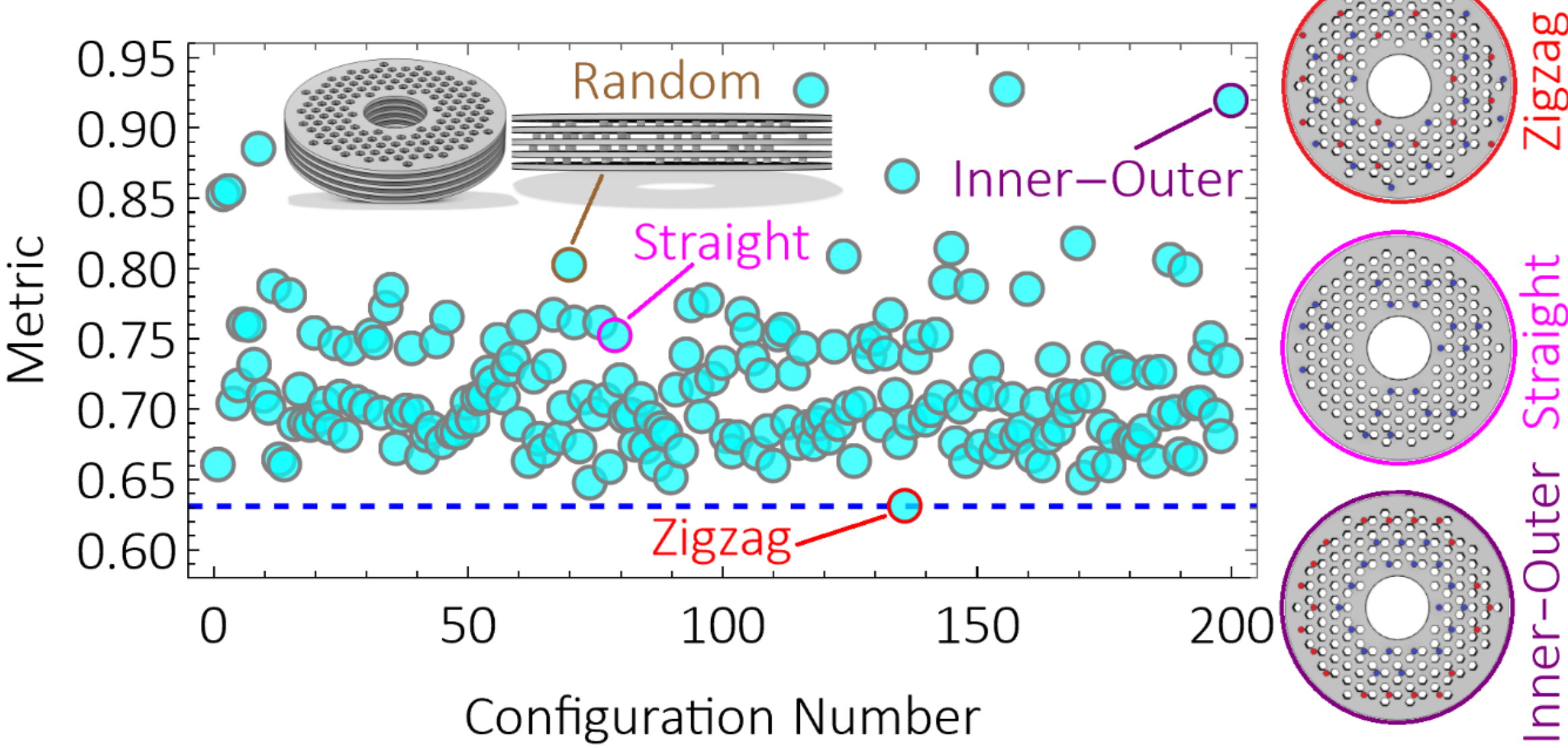

**Fig. 2. Thermal and mechanical characteristics of layered metamaterials.** Effect of pillar locations (20 pillars per layer interface) on thermal conductivity and mechanical rigidity in a 5-layer system. The combined performance metric is defined as $[(\kappa_{\mathrm{eff}}/\kappa_0)^2/2+(\sigma_{\max}/\sigma_0)^2/2]^{(1/2)}$. Red and blue dots in the right panel indicate the positions of the pillars on the top and bottom surfaces of each layer, respectively. The outer diameter, inner diameter, and thickness of each layer are 34.9 mm, 10 mm, and 0.6 mm, respectively. Each pillar has a height of 0.7 mm and a diameter of 0.9 mm. Hexagon-to-hexagon center spacing and hexagon apothem are 1.3 mm and 0.435 mm, respectively, for the hexagonal voids in each layer.

We also simulated the "straight" pillar configurations, where the pillars above and below each layer are vertically aligned (middle-right panel in **Fig. 2**). Our simulations of 30 different straight pillar configurations all yield higher combined metrics than those of the four possible different zigzag configurations, highlighting the latter configurations' potential as highly efficient pillar arrangements for optimized mechanical and thermal properties. This unique performance of the zigzag configurations can be attributed to the rigid structure of each layer, supported by well-distributed interlayer pillars, leading to a distinct distribution of heat and mechanical stress within and across the zigzag layered structures.

### 2.3. Thermal Optimization of Zigzag Layered Structures

Having established the overall optimal balance between the thermal conductivity and mechanical rigidity of the zigzag configurations, we now focus on optimizing their $\kappa_{eff}$ by varying the number and thickness of layers and pillars (**Fig. 3**). For identical geometry but varying numbers of layers, we observe that $\kappa_{eff}$ decreases by more than twofold as the number of layers increases from 2 to 16, suggesting that layered structures can achieve progressively superior thermal performance compared to their solid counterparts with increasing the number of layers. Similarly, in the 16-layer system, the zigzag configuration demonstrates a 30% reduction in $\kappa_{eff}$ compared to the straight configuration, contrasting with a minor variation in the thermal performance between AB and ABC stacking modes at the unit cell level. This highlights the importance of utilizing a delicately designed thermal insulating unit cell as the fundamental building block, combined with repetitions of zigzag stacking of several layers, to obtain high-performance thermal metamaterial structures.

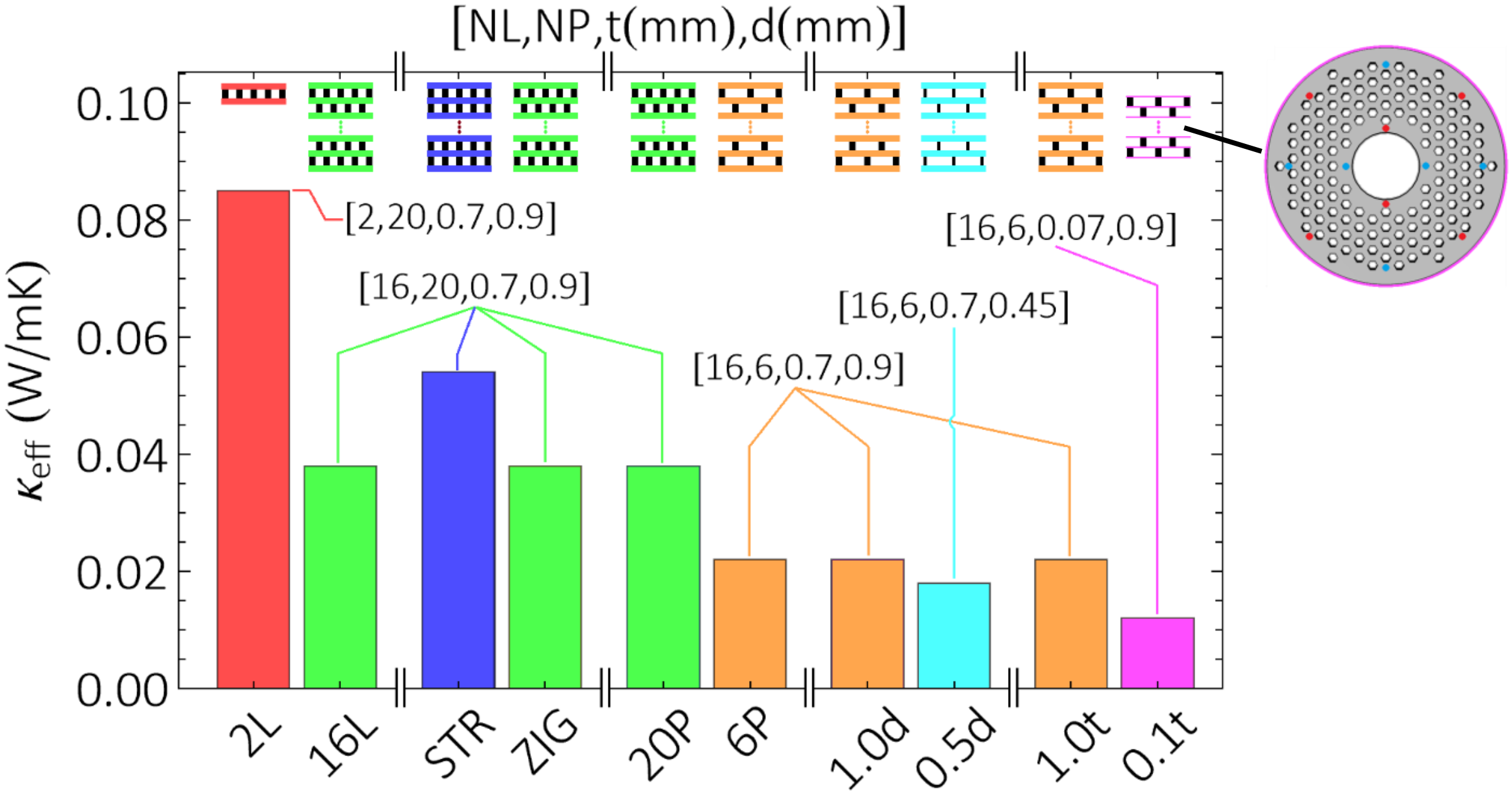

**Fig. 3. Optimization of thermal properties of zigzag layered structures.** Effect of the number of layers (NL), number of pillars (NP), layer thickness (t, mm) and pillar diameter (d, mm) on $\kappa_{eff}$ of layered systems. Red and blue dots in the right panel indicate the positions of the pillars on the top and bottom surfaces of each layer, respectively, in the 16-layer system with the 6-pillar zigzag configuration. The outer and inner diameters of each layer are 34.9 mm and 10 mm, respectively. Each pillar has a height of 0.7 mm. The hexagon-to-hexagon center spacing and the apothem of the hexagonal voids in each layer are 1.3 mm and 0.435 mm, respectively.

Within the 16-layer zigzag pillar system, decreasing the number of interlayer pillars from 20 to 6 results in more than a 40% reduction in $\kappa_{\text{eff}}$. This improvement is notable, considering that the total volume fraction of the system decreases by less than 1%, while the overall interface area between the layers and pillars is reduced by 70%. Similarly, halving the pillar diameter in the 16-layer system with the 6-pillar zigzag configuration reduces the total volume fraction by less than 0.3% but results in a reduction of 18% in $\kappa_{\text{eff}}$. This indicates that both the number and diameter of the pillars can significantly reduce $\kappa_{\text{eff}}$ with minimal impact on the total volume fraction of the layered structures.

Finally, decreasing the layer thickness by an order of magnitude in the 16-layer system with the 6-pillar zigzag configuration achieves an over 45% reduction in $\kappa_{\text{eff}}$, reaching an ultra-low value of 12.5 mW/(*m.K*). This performance theoretically exceeds the best-reported thermal insulating ceramic aerogels, such as zircon nanofiber-based aerogels, which exhibit a thermal conductivity of 58 mW/(*m.K*) at 500 °C **[14]**, highlighting the superior thermal insulation performance of well-engineered layered structures compared to randomly twined fibrous aerogel structures. However, despite the 45% improvement in $\kappa_{\text{eff}}$, the 16-layer system with a layer thickness of 0.07 mm exhibits mechanical rigidity that is approximately two orders of magnitude weaker compared to its counterpart with a layer thickness of 0.7 mm, emphasizing the need for carefully engineered configurations to address specific application requirements.

## 3. CONCLUSIONS

This study explores the potential of a novel class of layered metamaterials, inspired by 2D crystal architectures, to achieve a notable balance between ultra-low thermal conductivity and high mechanical rigidity, addressing the limitations of conventional materials. Our finite element simulations predict that zigzag configurations can theoretically achieve an ultra-low thermal conductivity of 12.5 mW/(*m.K*), while also maintaining robust mechanical stability. The reconfigurable nature of these structures, enabled by tunable parameters such as the number, thickness, and placement of layers and pillars, allows for customizable performance. Strategies like reducing interlayer pillars, decreasing pillar diameters, and minimizing layer thickness substantially enhance thermal insulation with minimal impact on volume fraction. These findings emphasize the paradigm-shifting potential of well-engineered layered metamaterials for applications in extreme environments, lightweight aerospace structures, and advanced thermal management systems, paving the way for next-generation super-insulating materials with tailored mechanical properties.

## ACKNOWLEDGEMENTS

This work was supported by the National Science Foundation (NSF CNS-2019131). The application of our designs in space engineering was supported by an Early Career Faculty grant from NASA's Space Technology Research Grants Program (Grant no. 80NSSC25K7832).

# Supporting Information

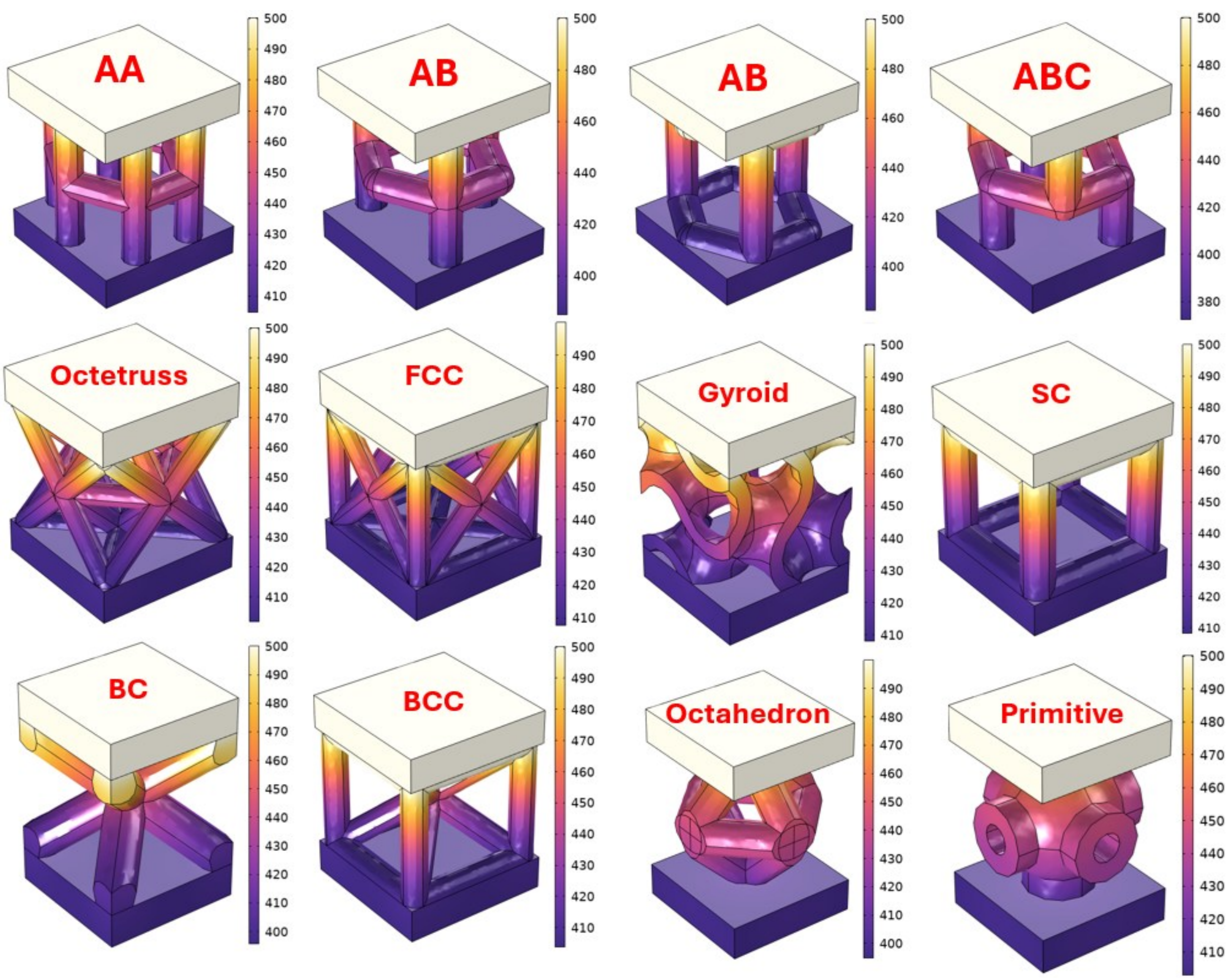

**Fig. S1.** Finite element temperature distributions for single 2 mm cubic zirconia unit cells with a volume fraction of 32%, sandwiched between solid plates. The top plate is maintained at 500 °C. Temperature values are shown in degrees Celsius. Note that the AB unit cell can adopt two different configurations, but both yield identical temperature distributions.

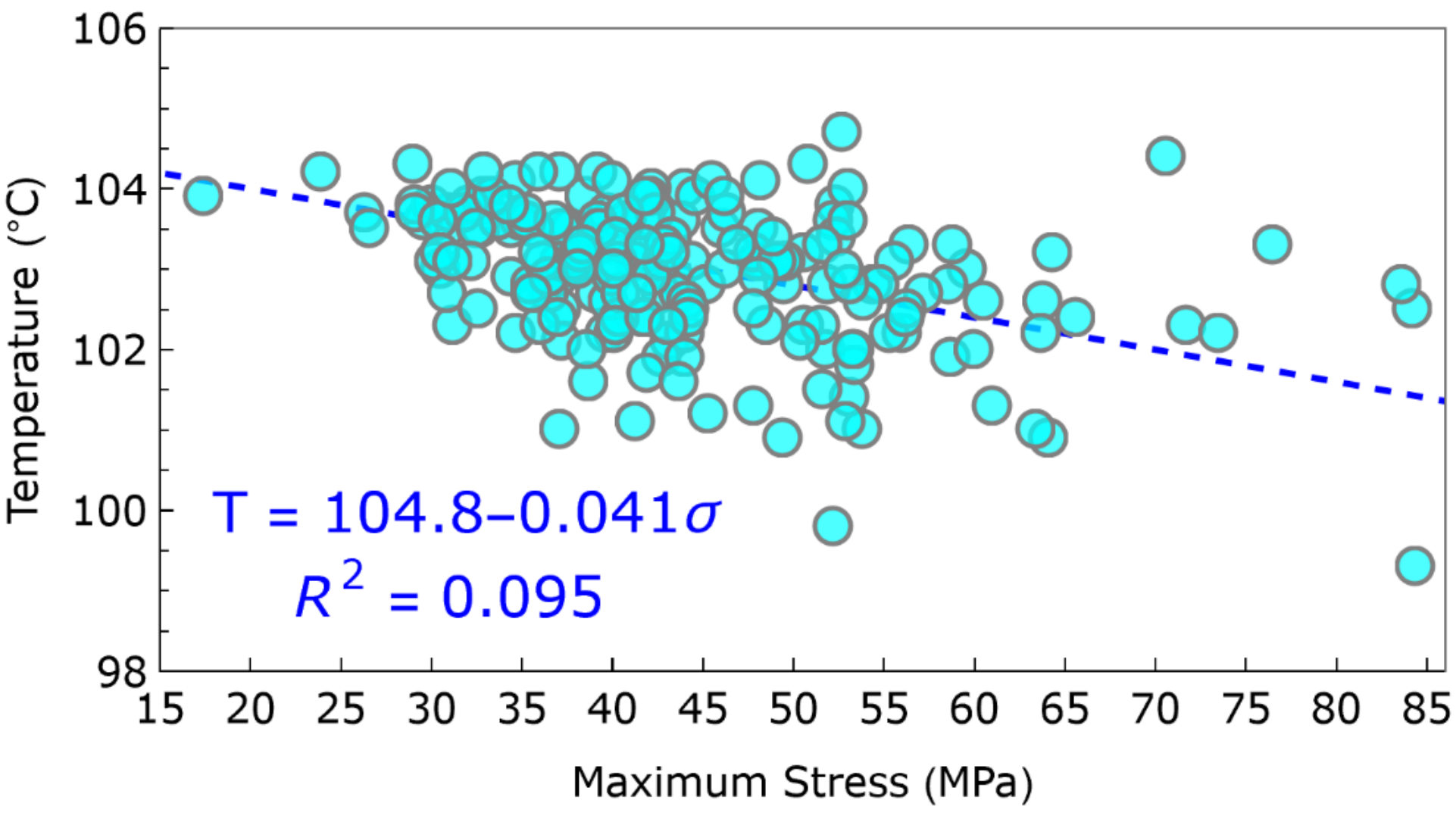

**Fig. S2.** Temperature of the bottom-most layer versus maximum stress within layered structures under a 10 N load, illustrating a weak correlation between thermal conductivity and mechanical rigidity.